# Conductivity enhancement in plastic-crystalline solid-state electrolytes

K. Geirhos[1], P. Lunkenheimer[1]*, M. Michl[1], D. Reuter[1] and A. Loidl[1]

**Finding new ionic conductors that enable significant advancements in the development of energy-storage devices is a challenging goal of current material science. Aside of material classes as ionic liquids or amorphous ion conductors, the so-called plastic crystals (PCs) have been shown to be good candidates[1,2,3,4] combining high conductivity and favourable mechanical properties. PCs are formed by molecules whose orientational degrees of freedom still fluctuate despite the material exhibits a well-defined crystalline lattice. Here we show that the conductivity of Li$^+$ ions in succinonitrile, the most prominent molecular PC electrolyte[3], can be enhanced by several decades when replacing part of the molecules in the crystalline lattice by larger ones. Dielectric spectroscopy reveals that this is accompanied by a stronger coupling of ionic and reorientational motions. These findings, which can be understood in terms of an optimised "revolving door" mechanism, open a new path towards the development of better solid-state electrolytes.**

Ionic conductors are essential for a number of applications, including energy-storage and -conversion devices as batteries, fuel cells, supercapacitors and solar cells. The most natural choice for such materials seem to be liquids that act as solvents for ionic species as Li$^+$ or Na$^+$ and provide a highly mobile host system enabling fast ionic motions. Indeed, even in modern electrochemical devices as, e.g., the Li-ion battery, liquid electrolytes are still commonly used. Nevertheless, various strategies are pursued to achieve further advances in electrolyte technology. One is the development of superior liquid materials, the most prominent current example being the so-called ionic liquids[5]. An alternative approach is the use of *solid* materials with high ionic conductivity, thus avoiding the common shortcomings of liquid electrolytes, e.g., leakage and flammability. Materials with special crystal structures allowing for sizable ionic diffusion, at least at high temperatures, are known since long (e.g., *β*-alumina[6]). Amorphous solids are another interesting option, e.g., polymers as polyethylene oxide[7,8]. The most recent development are plastic crystals, some of which have been shown to reach technically relevant conductivity values of > $10^{-4}$ $\Omega^{-1}$cm$^{-1}$ around room temperature when ions are added[1,2,3,4,9,10,11].

In plastic crystals, the centres of mass of the molecules are fixed on a crystalline lattice with translational symmetry. However, their orientational degrees of freedom more or less freely fluctuate at high temperatures and often show glassy freezing at low temperatures[12]. The molecules of most PCs have roughly globular shape and relatively weak mutual interactions, providing little hindrance for reorientational processes. This often leads to high plasticity, thus explaining the term "plastic crystal". This plasticity can be beneficial for application, enabling adaption to mechanical stresses. Ionic charge transport in PCs is often thought to be directly coupled to these reorientational motions via a "revolving door" or "paddle wheel" mechanism[2,9,13,14,15]: The reorientations may generate transient free volume within the lattice enabling ionic motion. Various types of defects in the PC phase were also considered to explain the high ionic mobility[2,3,11,16,17,18,19]. Two classes of ionically conducting PCs are known, "classical" PCs consisting of a single neutral molecular species[3,10,11] and ionic PCs, composed of at least two types of ions[1,2,4,9,13,16,18] (some of the latter in fact can be regarded as ionic liquids if considering the usual definition of this material class[5] comprising a melting point of less than 100 °C). In addition to the charge transport of the dopant ions like Li$^+$ or Na$^+$, which is prerequisite for many electrochemical applications, ionic PCs can also exhibit some self-diffusion of the ions constituting the host material[2,18], which may hamper their applicability.

The best known ionically conducting, but non-ionic PC is succinonitrile (SN), C$_2$H$_4$(CN)$_2$. In two pioneering works by Long *et al.*[10] and Alarco *et al.*[3], it was demonstrated that the conductivity of SN around room-temperature reaches values up to $10^{-3}$ - $10^{-2}$ $\Omega^{-1}$cm$^{-1}$ when adding ions by admixing several percent of selected salts like LiTFSI (lithium *bis*-trifluoromethanesulphonimide). Even pure SN exhibits non-negligible ionic conductivity, most likely due to small amounts of impurities[10,20]. A lot of effort has been made to optimise the conductivity in SN by varying the type and amount of added salt[3,10,11,21] and SN also was suggested to be used in composites with polymer electrolytes, leading to an enhanced conductivity[22,23].

The plastic phase of SN is limited to a rather small temperature range of about 235 - 330 K. In contrast to many other PCs[12], it is difficult to supercool and at *T* < 235 K the material transfers into an orientationally ordered crystal. However, the addition of the related molecular compound glutaronitrile (GN), C$_3$H$_6$(CN)$_2$ strongly extends the plastic-crystalline region, enabling the transition into a so-called glassy crystal[20,24]. The glassy freezing of the orientational degrees of freedom in this system was investigated in detail[20,24]. In the present work we report that, unexpectedly, the addition of GN to SN doped with Li salt leads to a marked enhancement of the ionic conductivity by up to three decades. In contrast to varying the added ionic compound[3,10,11,21], the partial substitution of the PC molecules also seems to be a promising approach to optimize plastic-crystalline solid-state electrolytes.

To check for the influence of GN on the ionic conductivity, we investigated several SN$_{1-x}$GN$_x$ mixtures with *x* between 0 and 0.8. As conducting ion we used Li$^+$, introduced into the mixtures by adding 1% LiPF$_6$, a commonly used salt for electrolytes. In contrast to salts as LiTFSI, its anion has no dipolar moment, which facilitates the interpretation of dielectric-spectroscopy results. In the Supplementary Information, DSC measurements are provided revealing the melting points and other transitions in the investigated mixtures. The dc conductivity and the relaxation times characterizing the reorientational motions in the SN-GN mixtures were determined from dielectric measurements. A typical example is presented in Fig. 1 for $SN_{0.6}GN_{0.4}$ showing the frequency and temperature-dependent dielectric loss $\varepsilon''$. The two dominating microscopic processes in conducting PCs, ionic charge transport and molecular reorientation, are nicely revealed by the red and blue/green regions of this plot, respectively. Via the relation $\varepsilon''_{dc} = \sigma_{dc} / (\varepsilon_0 \omega)$ ($\varepsilon_0$ is the permittivity of free space and $\omega = 2\pi\nu$), the ionic dc conductivity of the sample leads to the low-frequency divergence of $\varepsilon''(\nu)$ observed at the higher temperatures. The blue and green curves, revealing loss peaks, signify a typical relaxational process[20,24]. Its continuous shift to lower frequencies with decreasing temperature mirrors the glassy freezing of the orientational degrees of freedom as also observed for the ion-free $SN_{1-x}GN_x$ mixtures[20,24]. At the lowest temperatures, indications for a faster, secondary relaxation process are found. It was treated in detail in ref. 20.

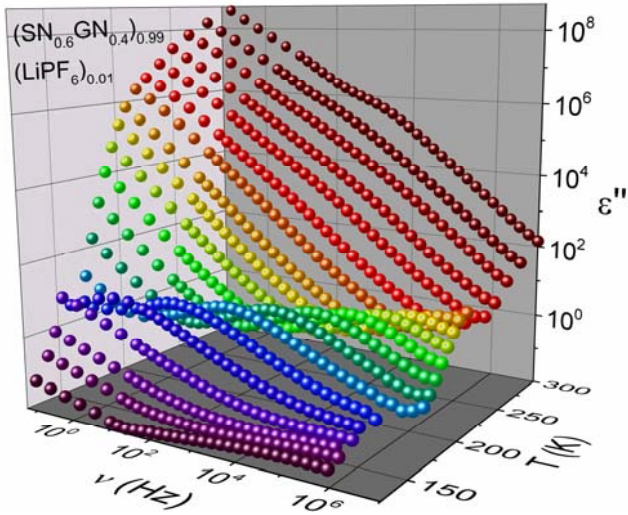

**Figure 1 | Dielectric loss spectra revealing the two microscopic processes in plastic crystals.** The figure shows spectra for $SN_{1-x}GN_x$ mixtures with 1% $LiPF_6$ and $x = 0.4$ as measured at different temperatures. While the red curves are dominated by ionic charge transport, the blue and green curves signify molecular reorientation processes.

Figure 2a shows the frequency-dependent dielectric constant $\varepsilon'$ as measured at various temperatures. The steplike decrease observed at high frequencies for 149 - 188 K arises from the molecular reorientations. In contrast to the nominally ion-free SN-GN system[20,24], the addition of 1% Li salt leads to a huge increase of $\varepsilon'(\nu)$ at low frequencies, reaching values of up to $10^8$. This finding can be ascribed to electrode polarization or blocking electrodes and is a typical phenomenon of ionic conductors[25]. It arises from the simple fact that the ions cannot penetrate into the metallic capacitor plates.

The intrinsic dc conductivity of the sample corresponds to a plateau in $\sigma'(\nu)$ (Fig. 2b). The electrode blocking causes the observed reduction of $\sigma'$ at low frequencies, which is most pronounced at the highest temperatures due to the high ionic mobility. The electrode effects in this system seem to occur in two successive steps, leading to an s-shaped curvature in the spectra at low frequencies and high temperatures. Similar behaviour is often found in ionic conductors investigated in a sufficiently broad frequency/temperature range and possible microscopic origins were discussed, e.g., in ref. 25. In ref. 26 similar effects, observed in SN doped with Li salts, were interpreted as indications for an intrinsic relaxation process arising from the trans-gauche isomerism of the SN molecules. As this led to unrealistically high relaxation strengths of the order of $10^6$, to us electrode polarization seems the most likely explanation.

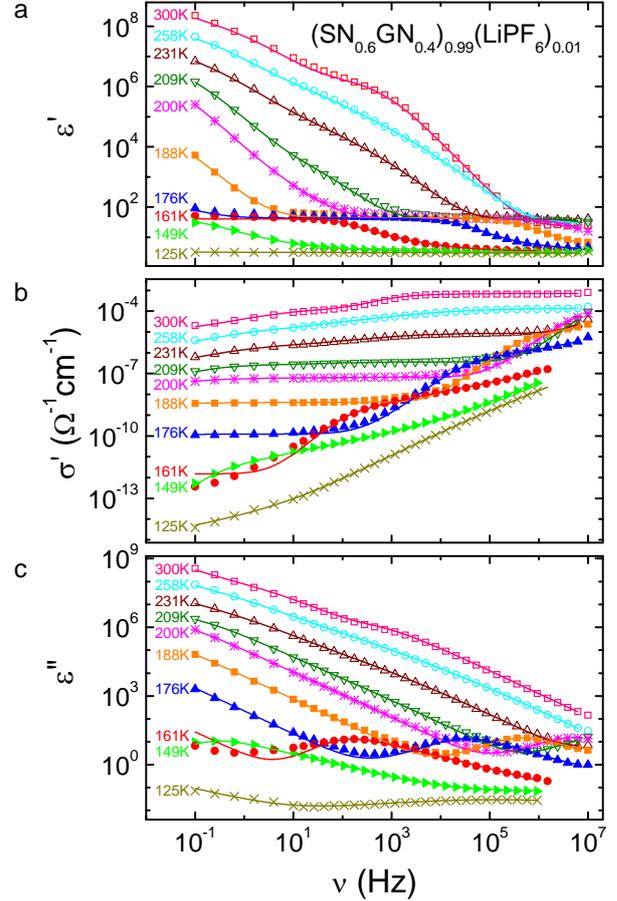

**Figure 2 | Typical dielectric and conductivity spectra of a $SN_{1-x}GN_x$ mixture with 1% $LiPF_6$.** Spectra of the dielectric constant (**a**), conductivity (**b**) and loss (**c**) are shown for $x = 0.4$ and various temperatures. The lines in **a** and **c** are fits assuming two distributed RC circuits to account for the blocking electrodes, a dc-conductivity contribution and two relaxation functions for the reorientational motions as explained in the text. The lines in **b** are calculated from the fit curves in **c** via $\sigma' = \varepsilon'' \, \varepsilon_0 \, \omega$.

Accounting for the different contributions discussed above, simultaneous fits of the spectra of $\varepsilon'$ and $\varepsilon''$ were performed (lines in Fig. 2). The blocking electrodes were modelled by two distributed RC circuits, connected in series to the bulk contribution[25]. For the dc conductivity $\sigma_{dc}$ a term $\varepsilon''_{dc} = \sigma_{dc} / (\varepsilon_0 \omega)$ was introduced. The relaxations were fitted by the empirical Cole-Davidson function (main relaxation) or the Cole-Cole function (secondary relaxation), a common approach in glass physics[12,27]. Overall, the achieved fits are excellent, except for small deviations around the minimum in $\varepsilon''(\nu)$ at the left wing of the main relaxation peak (Fig. 2c). In ref. 20 these deviations, also seen in the ion-free system, were discussed in terms of possible ac conductivity or an additional relaxation process, which, however, is out of the scope of the present work. The determination of $\sigma_{dc}$ at low temperatures is hampered by this effect and thus only data at $T > 170$ K are used for the further evaluation. The other investigated SN-GN mixtures showed qualitatively similar spectra as in Figs. 1 and 2.

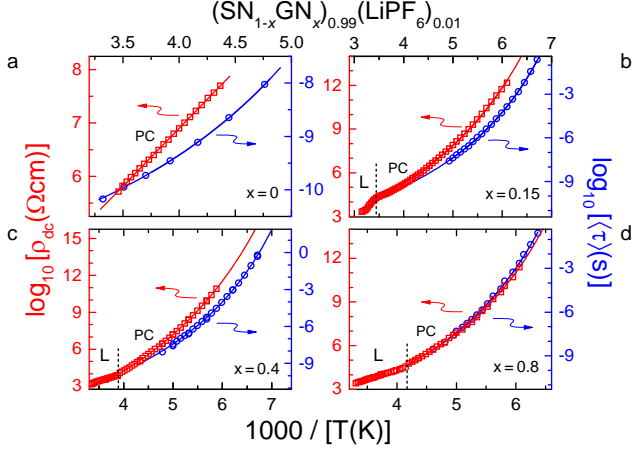

**Figure 3 | Temperature dependence of the ionic and reorientational dynamics in SN$_{1-x}$GN$_x$ mixtures with 1% LiPF$_6$.** The dc resistivity (squares, left scale) and reorientational relaxation times (circles, right scale) as deduced from the dielectric spectra are shown in Arrhenius representation. (The provided average relaxation times $\langle \tau \rangle$ were determined from the parameters of the CD function[20,24]. For $x = 0$, the $\tau$ data represent the behaviour for SN without LiPF$_6$.) Data for pure SN without GN (**a**) and three SN-GN mixtures (**b** - **d**) are shown. In each frame, the left and right ordinates were scaled to achieve the same number of decades per scale unit. Obviously, GN addition increases the coupling of the translational ionic and reorientational molecular motions. The solid lines are fits with the VFT law. For $x = 0$, $\rho_{dc}(T)$ could be fitted by an Arrhenius law. The vertical dashed lines indicate the melting points, in reasonable agreement with the DSC measurements provided in the Supplementary Information.

Figure 3 shows the temperature dependence of the two most important parameters, dc resistivity $\rho_{dc} = 1/\sigma_{dc}$ (left scale) and reorientational relaxation time $\tau$ (right scale) as determined from the spectra. The data are provided in Arrhenius representation for pure SN and three SN-GN mixtures. To enable a comparison, the $\rho_{dc}$ and $\tau$ ordinates are scaled to achieve the same number of decades per scale unit. Starting values of the ordinates were chosen to make the $\rho_{dc}$ and $\tau$ curves match at the highest investigated temperatures. Without GN addition (Fig. 3a), charge transport and reorientational motions are clearly decoupled: While $\rho_{dc}$ follows thermally activated behaviour, $\rho_{dc} \propto \exp[E/(k_B T)]$, $\tau(T)$ can be better described by the empirical Vogel-Fulcher-Tammann (VFT) law known from glass physics[27], usually written in the modified form[28]:

$$\tau = \tau_0 \exp\left[\frac{D T_{VF}}{T - T_{VF}}\right] \qquad (1)$$

Here $D$ is the so-called strength parameter[28], $\tau_0$ an inverse attempt frequency and $T_{VF}$ is the Vogel-Fulcher temperature, where $\tau$ diverges. For $x > 0$, VFT behaviour is also found for the resistivity. For the highest concentration, $x = 0.8$, the $\tau(T)$ and

$\rho(T)$ curves can perfectly be scaled onto each other (Fig. 3d). Overall, the addition of GN to SN seems to increase the coupling of the translational ionic and reorientational molecular motions. To understand this finding, it seems reasonable to assume that reorientations of the bulkier, more stretched GN molecules expand the lattice and are more effective in opening paths for translational motions of the ions. For high GN concentrations, sufficient numbers of these highly effective "revolving doors" are available and this mechanism seems to dominate the ionic charge transport (however, pure GN cannot be used as it has no plastic phase[20,24]. Alternatively, a more defect-dominated mechanism also seems possible as, of course, admixing GN implies an increase of the defect concentration.

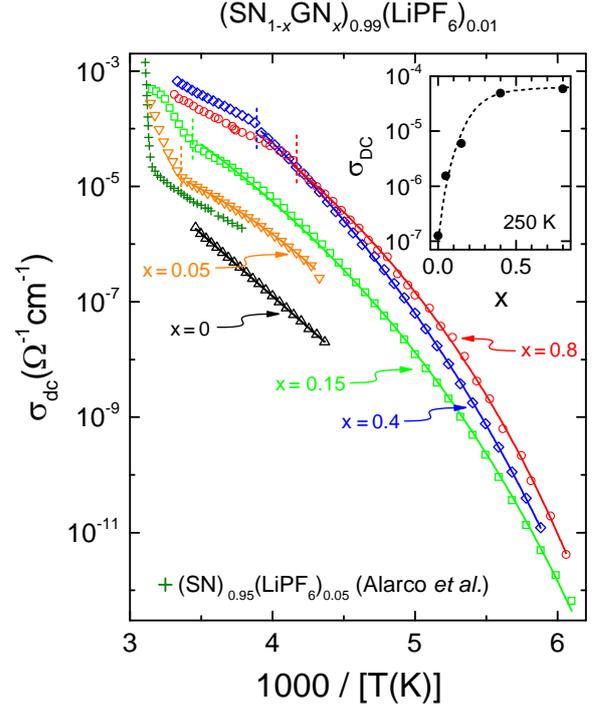

**Figure 4 | Influence of GN content on the temperature-dependent ionic conductivity in SN$_{1-x}$GN$_x$ mixtures with 1% LiPF$_6$.** The dc conductivity is shown in Arrhenius representation for all investigated mixtures. The solid lines are fits with the VFT equation, eq. (1), modified for the conductivity: $\sigma_{dc} = \sigma_0 \exp[-DT_{VF}/(T-T_{VF})]$. The vertical dashed lines indicate the melting points (cf. DSC measurements in the Supplementary Information). The plusses show the results for pure SN, doped with 5% LiPF$_6$ as published in ref. 3. The inset shows the dependence of $\sigma_{dc}$ in the plastic phase on SN concentration. Obviously, adding GN strongly enhances the ionic conductivity of plastic-crystalline SN.

Does the rising importance of this mechanism when GN is added also lead to more effective charge transport, reflected in an increase of conductivity? This is clarified in Fig. 4, comparing the dc conductivities of the investigated mixtures. Indeed, a huge increase of $\sigma_{dc}$ is found: Even for the smallest GN concentration of only 0.05 (inverted triangles) it becomes enhanced by more than one order of magnitude compared to pure SN (upright triangles; see also inset of Fig. 4, where $\sigma_{dc}(x)$ at 250 K is shown). Despite the salt concentration is only 1%, the conductivity of this mixture exceeds that of pure SN with 5% LiPF$_6$ as reported in ref. 3 (plusses in Fig. 4). Obviously, 5% GN addition leads to a marked improvement of ionic *mobility*. This generates a stronger increase of conductivity than the fivefold enlargement of the ion *density* when adding 5% $LiPF_6$ to pure SN (plusses in Fig. 4) instead of 1% (upright triangles). Moreover, Fig. 4 demonstrates that, when adding larger amounts of GN, the conductivity of SN even can be improved by up to three decades. Saturation of $\sigma_{dc}(x)$ is approached above $x \approx 0.4$ (inset of Fig. 4).

Overall, our detailed dielectric investigations reveal a marked increase of ionic conductivity when GN is added to the prominent solid-state electrolyte SN, which is accompanied by a stronger coupling of molecular reorientation and ionic translation. It seems that an effective revolving-door mechanism achieved by adding the larger GN molecules to SN is the key to considerably enhance ionic mobility and, thus, conductivity. However, the role of possible defect mechanisms and of the trans-gauche isomerism of the involved molecules remains to be clarified. It should be noted that the melting point of $SN_{1-x}GN_x$ mixtures decreases with GN concentration[20,24]. Therefore only for $x = 0.05$, the samples are in the PC state at room temperature. While is was not the aim of the present work to find a new PC ionic conductor optimized for application, our findings demonstrate a new path to a considerable conductivity enhancement in PCs. To find solid-state electrolytes, optimally suited for application, performing further experimental work following this path, e.g., by varying the added salt and exploring the feasibility for other PCs, seems a promising approach.

## Methods

**Sample preparation.** SN and GN with purities > 99% were purchased from Arcos Organics. $LiPF_6$ (purity 99.99%) was obtained from Sigma-Aldrich. Appropriate amounts of GN (liquid at room temperature) and $LiPF_6$ powder were added to SN powder[24]. For low $x$, the mixtures were slightly heated to achieve complete dissolution. The samples were characterized by differential scanning calorimetry (DSC) (see Supplementary Information).

**Dielectric measurements.** The dielectric measurements were performed using a frequency-response analyser (Novocontrol Alpha-analyser). The sample materials were filled into parallel-plate capacitors with plate distances of 0.1 mm. For cooling and heating of the samples, a $N_2$-gas cryostat was used. The measurements for $x \leq 0.4$ were done under heating. To avoid ambiguities arising from the complex succession of different phases known to occur at higher SN concentrations under heating[24], the measurements for $x = 0.8$ were done under cooling (see also Supplementary Information).

## Acknowledgements

This work was supported by the Deutsche Forschungsgemeinschaft via Research Unit FOR1394.

## Author contributions

A.L. and P.L. conceived and supervised the project. K.G. and M.M. performed the dielectric measurements. D.R. performed the DSC measurements. K.G. analysed the data. P.L. wrote the paper. All authors discussed the results and commented on the manuscript.

## Additional information

Correspondence and requests for materials should be addressed to P.L.

# Conductivity enhancement in plastic-crystalline solid-state electrolytes

## Supplementary Information

**K. Geirhos[1], P. Lunkenheimer[1]\*, M. Michl[1], D. Reuter[1] and A. Loidl[1]**

[1]Experimental Physics V, Center for Electronic Correlations and Magnetism, University of Augsburg, 86159 Augsburg, Germany

\*e-mail: peter.lunkenheimer@physik.uni-augsburg.de

## 1. DSC measurements

Figure S1 shows the DSC heat flow (HF) as measured under cooling and heating for various $SN_{1-x}GN_x$ mixtures with 1% $LiPF_6$. For $x = 0.05 - 0.4$ (frames a - c), the measurements were performed with a rate of 10 K/min. For $x = 0.05$, the negative (i.e., exothermic) peak, observed under cooling at about 320 K, indicates the transition into the plastic crystal (PC). Below about 190 K, the sample transforms into an orientationally ordered crystal, revealed by a second exothermic peak. Under heating, the endothermic peaks at about 235 and 325 K arise from the transitions into the plastic-crystalline and the liquid phases, respectively, in agreement with the phase diagram published in ref. 1 for $SN_{1-x}GN_x$ without salt admixture.

For $x = 0.15$ and 0.4 (Figs. S1b and c), the plastic phase reached at about 300 and 245 K, respectively, can be easily supercooled[24]. The observed low-temperature anomalies showing up around 147 K reveal the typical signature of glassy freezing, signifying the transition from the so-called glassy crystal with frozen orientational motions to the PC. Finally, the exothermic peaks detected at about 310 ($x = 0.15$) or 255 K ($x = 0.4$) under heating arises from the melting of the PC. Generally, the amplitude of the observed melting peaks diminishes when $x$ increases from 0.05 to 0.4. This can be ascribed to the stronger substitutional disorder for higher $x$, thus leading to a smaller entropy change when passing from the plastic-crystalline to the liquid state.

For $x = 0.8$ without salt addition, in ref. 1 a rather complex phase behaviour was revealed. Especially, it was demonstrated that at this concentration the material may also transfer into a supercooled-liquid state when cooled with a rate $\geq 5$ K/min. However, in our DSC experiments for samples with 1% $LiPF_6$, we found that this state was also attained with only moderate rates. For the measurement shown in Fig. S1d, we applied a slower cooling rate of 2 K/min trying to reach the plastic-crystalline state. However, in contrast to the results for smaller $x$, under cooling no indication of the transition into this state was found (it should occur below 230 K and thus the limited maximum temperature of about 275 K in Fig. S1d represents no problem). Only under heating, a small melting peak at about 230 K (in agreement with ref. 1) indicates that at least part of the sample has transferred into the PC state.

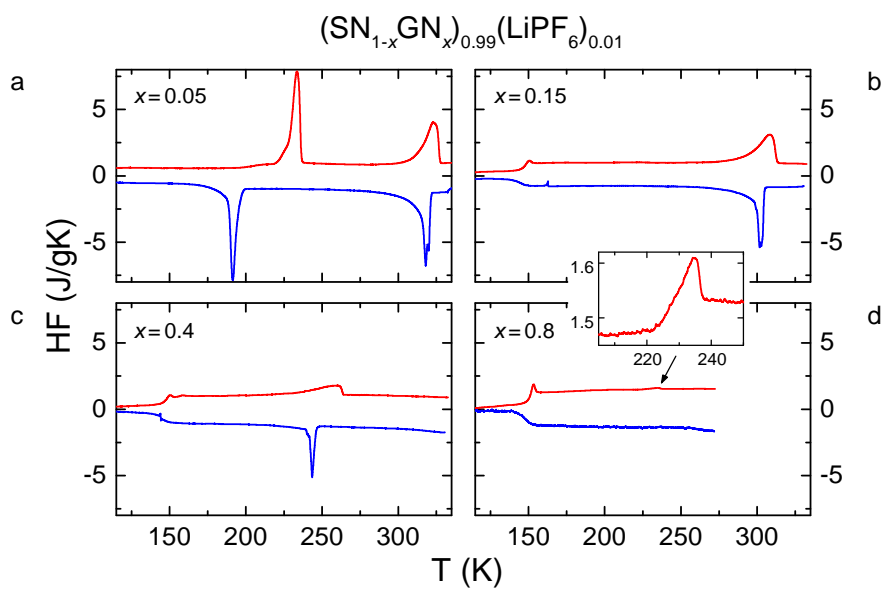

**Figure S1 | DSC curves for $SN_{1-x}GN_x$ mixtures with 1% $LiPF_6$, measured under cooling (blue) and heating (red).** To enable a comparison of the different curves, the heat flow per sample mass is shown, divided by the cooling/heating rate.

## 2. Plastic-crystalline and supercooled-liquid states for *x* = 0.8

In contrast to the DSC experiments, for the dielectric measurements, performed with about 0.4 K/min, the plastic-crystalline state was clearly reached. This is demonstrated, e.g., by Fig. S2, which shows results on dc resistivity $\rho_{dc}$ and relaxation time $\tau$ and compares them for both the plastic-crystalline (circles and squares) and supercooled-liquid state (crosses and plusses). Both $\rho_{dc}$ and $\tau$ are significantly smaller in the latter phase, which seems a reasonable finding. To reach the supercooled-liquid phase, the sample was first quenched and measured under heating as described in ref. 1. Above about 170 K, crystallization occurred. Obviously, due to the different cooling/heating rates and sample environments (aluminium pans or parallel-plate brass capacitors, respectively), the phase transitions observed in the DSC and dielectric experiments do not agree for the *x* = 0.8 sample.

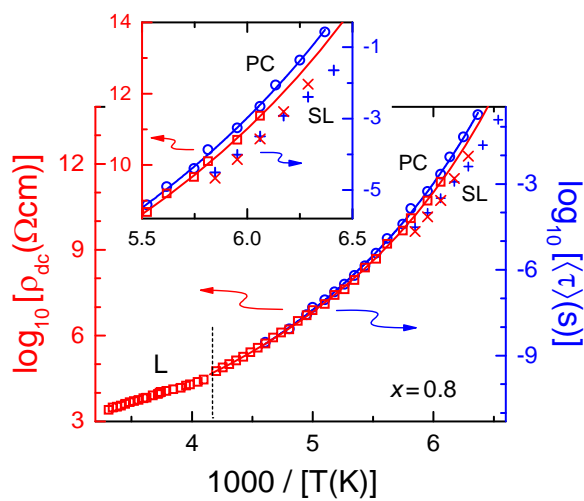

**Figure S2 | Temperature dependence of the ionic and reorientational dynamics in $SN_{0.2}GN_{0.8}$ with 1% $LiPF_6$.** The dc resistivity (squares and crosses, left scale) and reorientational relaxation times (circles and plusses, right scale) as deduced from the dielectric spectra are shown in Arrhenius representation. The squares and circles represent the results for the PC state (as shown in Fig. 2 of the main paper) and the crosses and plusses those for the supercooled liquid. The left and right ordinates were scaled to achieve the same number of decades per scale unit. The solid lines are fits with the Vogel-Fulcher-Tammann law (see main text). The vertical dashed line indicates the melting temperature. The inset shows a magnified view of the low-temperature region.